\documentstyle[12pt,aasms4]{article}
\def\reference{\par\noindent\hangindent=1cm\hangafter=1}

\newcommand{\eq}{\begin{equation}}
\newcommand{\ee}{\end{equation}}

\def\t0{\theta_{\circ}}

\def\be{\begin{equation}}
\def\en{\end{equation}}

\def\gapp{\ \lower 3pt\hbox{${\buildrel > \over \sim}$}\ }
\def\lapp{\ \lower 3pt\hbox{${\buildrel < \over \sim}$}\ }

\lefthead{Jiang, Laughlin, \& Lin}
\righthead{Brown Dwarfs}
\begin{document}

\title{On the Formation of Brown Dwarfs}

\author{Ing-Guey Jiang$^{1}$, G. Laughlin$^{2}$ and D.N.C. Lin$^{2}$}

\affil{$^1$ Institute of Astronomy, National Central University,
          Chung-Li, Taiwan}
\affil{$^2$ UCO/Lick Observatory, 
University of California, Santa Cruz, CA 95064, USA} 
 
\authoremail{jiang@astro.ncu.edu.tw}

\begin{abstract}
The observational properties of brown dwarfs
pose challenges to the theory of star formation.  Because their mass
is much smaller than the typical Jeans mass of interstellar clouds,
brown dwarfs are most likely formed through secondary fragmentation
processes, rather than through the direct collapse of a molecular
cloud core. In order to prevent substantial post-formation mass
accretion, young brown dwarfs must leave the high density formation
regions in which they form.  We propose here that brown dwarfs are
formed in the 
%optically thin outer regions of 
circumbinary disks.
Through post-formation dynamical interaction with their host binary
stars, young brown dwarfs are either scattered to large distance or
removed, with modest speed, from their cradles.
\end{abstract}


\section{Introduction}

Brown dwarfs are entities with mass below that require for hydrogen
burning to ignite ($< 0.075 M_\odot$) and above that associated with
gaseous giant planets ($\sim$ 13 Jupiter masses). 
%(Burrows et al 1997, 
%(Baraffe et al. 1998). 
Although the existence of brown dwarfs
was proposed by Kumar in 1963, their cool dim nature consigned them to
a strictly theoretical status for more than three decades.  Recently,
however, improved observational capabilities have led to the discovery
of many brown dwarfs, prompting a renaissance in our understanding of
these objects.
Brown dwarfs are now known as companions to main
sequence stars (Nakajima et al. 1995), as cluster members (Rebolo et
al. 1995,  Lucas \& Roche 2000)
as field objects (Ruiz et al. 1997), 
and as binary pairs (Basri \& Martin 1999).
Surveys
such as the Deep Near-Infrared Survey (DENIS) (Epchtein et al. 1994),
and the 2 Micron All-Sky Survey (2MASS) (Skrutskie et al. 1997), are
uncovering brown dwarfs in statistically significant quantities.  
%It now appears that brown dwarfs are as numerous, if not more numerous,
%than stellar mass objects
% (although they do not contribute
%significantly to the total mass of ordinary matter). 
%(Reid et al. 1999).

%Although brown dwarfs are now understood to be generally commonplace
%objects, they seem to be surprisingly rare as close companions to main
%sequence stars. 

A number of planet search programs based on the radial
velocity method have failed to uncover brown dwarfs orbiting in close
proximity (with separation less than $\sim 1$ AU) to stellar mass
primaries (Marcy \& Benitz 1989; Cochran \& Hatzes 1994; Mayor \&
Queloz 1995; Walker et al. 1995).  Brown dwarfs also appear to be rare
at intermediate separations ($1 < r < \sim$ a few AU) as evidenced by
the speckle imaging survey carried out by Henry \& McCarthy (1993).
Similarly, Oppenheimer et al. (2001) did a survey for companions of stars
within 8 pc of the Sun and found only one brown dwarf. 
 A small number of brown dwarfs have, however been found as binary
companions lying at separations much greater than a few AU 
(Neuh\"aeuser et al. 2000). 
Gizis et al. (2001) derived the wide companion frequency
and said the current data indicated that about one percent
of $M_{\rm V} < 9.5$ primaries have wide L dwarf companions;
the brown dwarf fraction should be 5-13 times higher.
To further complicate matters, 
brown-dwarf binary pairs 
(i.e. both components of the binary are brown dwarfs) have
been discovered. Basri \& Martin (1999) claimed that they 
detected the first spectroscopic brown-dwarf pair with a period of about 
5.8 days only. Koerner et al. (1999) found three brown-dwarf pairs
with the projected 
physical separations between 5 and 10 AU.
Recently, Gizis et al. (2003) presented analysis of HST images of 82 nearby
field late M and L dwarfs and estimated that binary fraction is about 
10 to 20 percent in the range 1.6--16 AU.

%%% These binary-brown-dwarf
%%%systems usually exhibit binary separation smaller than 40 AU
%%%(Basri \& Martin 1999).

%To summarize, these searches for brown dwarfs reveal that 
%(a) they are as numerous
%as M dwarfs both in the field and in young stellar clusters, (b) they
%are occasionally wide companions but rarely close companions to main
%sequence stars, and (c) brown dwarf-brown dwarf pairs tend to be close
%binary systems with separations less than a few AU.

Because the mass of brown dwarfs is much smaller than the
usual Jeans mass for a typical molecular cloud,
Lin et al. (1998) claimed that the encounter between two protostellar discs
might increase the Jeans mass locally.
Reipurth \& Clarke (2001) suggested that brown dwarfs are substellar objects 
because they have been ejected from newborn multiple systems. They use a simple
model of timescales to show that this could happen.
Bate et al. (2002) used a smoothed particle hydrodynamics code
to show that brown dwarfs could be formed  through the collapse and 
fragmentation of a turbulent molecular cloud multiply and thus confirmed
what was suggested by Reipurth \& Clarke (2001).

%brown dwarfs are of low mass because they are ejected ?
%(ejection might help to avoid getting more mass but no the only possibiliy
%. locally they might be settling down on a typical mass and stop increase
%without being really ejected (like long-period companion been observed)

%(like planet fomation by boss)

%brown dwarfs are ejected because they are of low mass ?
%Could be true but they have to be in unstable ``escape zone'' 

%Thus, we provide details study of this issue for simplified model.
 
%However, if all brown dwarfs have to be ejected to avoid accreting 
%too much mass, one might ask how it would be possible that they are 
%wide companions of solar-type stars sometimes.
In this paper,
we propose a formation scenario for brown dwarfs which provides a
natural explanation for the current observational situation.
Numerical simulations have confirmed that a rotating  protostellar cloud may
induce hierarchical fragmentation 
and the emergence of very low mass objects (Burkert \& Bodenheimer
1996). However, these entities continue to grow as they merge with
each other and accrete from the residual infalling envelope of the
cloud unless they are ejected from the dense central regions
(Bate et al. 2002). 
In many cases, the low-mass
objects become the binary companions of a dominant massive
protostellar object.  In such a binary configuration, gas in the
residual rotating infalling envelope settles onto a circumbinary disk
and is then preferentially accreted by the less massive member through
the outer Lagrangian point.  This tendency generally leads the
binaries to gradually acquire comparable masses.
Subsequently, mass accumulation in
circumbinary disk may lead to the
collapse of low mass entities and become candidates of brown dwarfs.

Therefore, it seems natural to suggest that perhaps many 
brown dwarfs could be formed through  fragmentation
of the circumbinary disks.
We study the ``escape zone'' where the brown dwarfs could be ejected
and become field stars.
We demonstrate that in some cases, the brown dwarfs get ejected
but in other cases, they could become long-period companion. 
We also study the criteria that the brown-dwarf pair can survive.
However, we do not intend to study the problem of 
brown dwarf desert here (please see the model by Armitage \& Bonnell 2002).
In Section 2, we present the results for single brown dwarf.   
The study about formation of brown-dwarf pairs would be in Section 3.
We provide concluding remarks in Section 4.

%Nevertheless, numerical simulations have confirmed that  
%rotation may
%induce hierarchical fragmentation 
%(Bodenheimer ?? Boss ?? Miyama ??)
%and the emergence of very low mass objects (Burkert \& Bodenheimer
%1996).  However, these entities continue to grow as they merge with
%each other and accrete from the residual infalling envelope of the
%cloud unless they are ejected from the dense central regions
%(Bodenheimer \& Burkert 1996 and others). In many cases, the low-mass
%objects become the binary companions of a dominant massive
%protostellar object.  In such a binary configuration, gas in the
%residual rotating infalling envelope settles onto a circumbinary disk
%and is then preferentially accreted by the less massive member through
%the outer Lagrangian point.  This tendency generally leads the
%binaries to gradually acquire comparable masses (Bate 1997, 2000).
%Subsequently, the tidal perturbation of binary stars with comparable
%masses induces efficient angular momentum transfer from the binary pair to
%the gas in the circumbinary disk, creates a gap, and imposes a barrier
%against inward gas diffusion (Artymowicz \& Lubow 1996, Lin \&
%Papaloizou 1993).

\section{Stability of brown dwarf companion around binary stars}

We now investigate the orbital stability and evolution of brown dwarfs
formed in circumbinary rings.  These companions are perturbed by the
tidal disturbance of the binary star's gravitational potential.
For computational convenience, we assume
these newly fragmented brown dwarfs quickly become centrally condensed
and the residual gas does not contribute significantly to the
gravitational potential such that the dynamics of the system may be
described by a few-body approximation.
A direct force integration of
the equation of motion is required for the computation of the orbital
evolution of this system.  We adopt a numerical scheme with Hermite
block-step integration which has been developed by Sverre Aarseth (Markino
\& Aarseth 1992, Aarseth, Lin \& Palmer 1993).

In this section, we further simplify the interaction procedure to a
three-body (the host binary star plus a brown dwarf) problem.  The
mass of each brown-dwarf fragment is sufficiently small that they do
not significantly perturb each other on the short term of
several thousand binary periods. 
%(this timescale corresponds to the
%persistent timescale of protostellar disks around typical T Tauri stars).  
In order to make a direct comparison with some existing
results, we first treat the brown dwarfs as massless particles. 
% This type of the three-body problem is in fact very important and
%fundamental for both planetary science and stellar dynamics and
%therefore it has been studied extensively.  (For example, Eggleton \&
%Kiseleva 1995, Laughlin \& Adams 1999, Holman \& Wiegert 1999 and
%references therein).  
But, in general, we carry out full 3-body
integration in which contribution due to the finite mass of the brown
dwarf is included.

We consider a range of ratio ($\mu = M_1 / (M_1 + M_2)$) of masses
($M_1$ and $M_2$) for the two components of the binary stars.
Following the approach by Holman \& Wiegert (1999), we consider, for
the host binary system, a range of orbital eccentricity ($e_\ast$).  
The semi major axis of the binary is set to be unity such
that all other length scales are scaled with its physical value. 
% Similar
% to the hydrodynamical simulations in the previous section, 
We also adopt $G (M_1 + M_2) =1$ such that the binary orbital period
is $2 \pi$. In this paper, the total mass of the central binary 
is assumed to be 1 $M_\odot$ in real 
unit.

Since we assume they are formed in circumbinary rings, 
we consider brown dwarfs with orbital semi major axis larger than that
of the binary system.  At the onset of the computation, 
all three stars are located at their apocenter
with respect to their common center of mass. 
% The results in \S2 show
It is possible 
that the fragments may have a range of orbital eccentricity ($e_b$).
Here, we consider two limiting eccentricity for the brown dwarf ($e_b
= 0$ and 0.4).  We also set the ratio ($\mu_b$) of brown dwarf's mass
to that of the binary to be 0.05 and we assume brown dwarfs rotate in
the same direction as the orbit of the binary secondary.

%\subsection{Fragments on Circular Orbits}
\subsection{Ejection criteria}
We choose a range of initial orbital semi major axis for the brown
dwarf for the $e_b =0$ case.  For each set of model parameters, we
adopt four values for the brown dwarf's angle of apocenter, 
%eccentric anomaly, 
$0^\circ$,
$90^{\circ}$, $180^{\circ}$ and $270^{\circ}$ with respect to that of
the binary system.

%In all our models, the orbital semi major axes of the brown dwarf is
%larger than that of the binary system.  
We are primarily seeking a
critical initial semi major axis ($a_c$), larger than which the brown
dwarf survives the binary system's perturbation within a timescale
$T_d$.  Our definition of survival is that the distance from the center of
mass of the system to the brown dwarf (starting with all four values of
the apocentric arguments) must
smaller than a critical value $R_d$. 
$R_d$ can be chosen to be a number which is much larger than the binary
 separation, say 10 or 100 times binary separation.
The choice of this number 
%might affect the results of critical semi major axis but 
would not affect the results much. Because as long as it is ejected, it will
go through both the boundary of 10 and 100 binary separation within $T_d$. 
If it is stable, it will stay within 10 binary separation always. 
Only marginally stable cases are 
more complicated, but these marginal cases only happen when initial
semi major axes 
are around particular values. 
Thus, we set
 the value of $R_d$ to be 25 binary separations.  In order to compare with
the results of Holman \& Wiegert (1999), we choose $T_d = 10^4$ binary period,
i.e. $T_d= 2 \pi \times 10^4$.
Based on several test runs, we find that the value of $a_c$ does not
change significantly if $T_d$ is increased to $10^6$ binary period. 
% That is
%brown dwarfs which can survive for $10^4$ yr can usually remain
%attached for a much longer timescale.  
Thus, we find $a_c$ to be a
useful parameter to classify our results.
% (Dvorak 2003).
% and their values are included in Figure 6--7.  
Although we use a totally different numerical scheme
as Holman \& Wiegert (1999), we are able to precisely reproduce the
results in Table 7 of their paper when we set the mass of brown dwarf
to be zero as they have.  But in general, we choose $\mu_b = 0.05$.
%\begin{table}
%\begin{center}
%\renewcommand{\arraystretch}{1.2}
%\begin{tabular}{lcc}
%\multicolumn{3}{c}{Table 1}\\
%\multicolumn{3}{c}{Critical Semimajor Axis for Circular Orbits($e_b=0$)}\\
%\hline
%        & $\mu=0.1$ & $\mu=0.5$\\
%\hline
%\hline
%e_\ast=0.0$ & 2.4 & 2.9\\
%e_\ast=0.2$ & 3.3 & 3.4\\	
%$e_\ast=0.4$ & 4.7 & 3.7\\
%$e_\ast=0.6$ & 6.0 & 3.7\\	
%\hline
%\end{tabular}
%\end{center}
%\end{table}

From these models, we find that the brown dwarf's ``escape zone'' (
with semi major axis $a < a_c$) is expanded slightly when they
have finite mass (see Fig.1 \& 2 for the comparison.)  The
expansion of the ``escape zone'' is larger for $\mu=0.1$ cases than
that for the $\mu=0.5$ cases because the motion of secondary star is
more affected by the finite mass of the brown dwarf. This effect is
particularly noticeable for the $\mu=0.1$ and $e_\ast=0.6$ case where $a_c$
is expanded from $3.9$ of massless particles to $6.0$ of brown-dwarfs
with $\mu_b =0.05$. The hydrodynamical simulations 
indicate that fragmentation of
circumbinary disks occurs primarily at around a few binary separation
away from their center of mass because this is the location where disk
gas may accumulate as a consequence of binary star's tidal torque.
Observations by Beckwith et al. (1990) 
indicated that about 42 percent of young stars
have discs. These discs' survival times are from $10^6$ to $10^7$
years and masses range from $10^{-3}$ to 1 $M_{\odot}$.
For example, Roddier et al. (1996) 
showed that the GG Tauri system has a circumbinary ring with the density 
peak at radius
about 2.7 times the semi major axis of the central binary orbit. The 
average density of this circumbinary ring is only about 
0.1 to 0.2 of the peak value from 60 AU to 160 AU,
then increases  rather sharply with a constant slope 
to reach the peak at 250 AU and finally, 
decays roughly as a power law, $R^{-2}$,  
until 460 AU from the center (see Fig. 9 of their paper). 
%The    ...2.7 .
%2.7 time of binary separation
%density :  
%peak around 2.7 binary separation  and  decay as R^-2 approximately,
%in general, fragment so , very compleciated,
%total mass:   about 0.1 of central stars for a protostellar system (Boss).
%total mass of central binary is 1 Solar mass %}
Thus, {\it most of the low-mass fragments formed in the circumbinary
rings have a high probability of being ejected by the gravitational
perturbation of their host binary systems.}

%\subsection{Fragments on Eccentric Orbits}
In general, the fragments might not  move on circular orbits. 
It would be interesting to consider the models in which the brown dwarfs
are assumed to move on eccentric orbits initially.
Since the circumbinary ring of GG Tauri is actually an ellipse with 
the eccentricity about $0.2$ (Roddier et al. 1996) and 
the fragments formed there might not exactly move along this ellipse,  
these fragments' initial orbital eccentricities could be even larger.

Therefore, 
we now consider a series of models with $e_b = 0.4$ while all other
parameters are similar to those for the $e_b=0$ case. 
% For each set of
%model parameters, we adopt four values for the brown dwarf's
%argument of the apogee, $0^\circ$, $90^{\circ}$, $180^{\circ}$,
%and $270^{\circ}$  with respect to that of the binary star.
%We determine the value of $a_c$ for different binary mass ratio $\mu$
%and $e_\ast$ in Fig. 2.
% and Figure 6--7.
The increases in $a_c$ in Fig. 1 and Fig. 2 
%compared with those in Fig. 1
clearly indicate that brown dwarfs with eccentric orbits are
definitely less stable than those with circular orbits.
%\begin{table}
%\begin{center}
%\renewcommand{\arraystretch}{1.2}
%\begin{tabular}{lcc}
%\multicolumn{3}{c}{Table 2}\\
%\multicolumn{3}{c}{Critical Semimajor Axis for Eccentric Orbits($e_b=0.4$)}\\
%\hline
%        & $\mu=0.1$ & $\mu=0.5$\\
%\hline
%\hline
%$e_\ast=0.0$ & 5.0 & 5.2\\
%$e_\ast=0.2$ & 5.5 & 6.4\\	
%$e_\ast=0.4$ & 6.6 & 6.8\\
%$e_\ast=0.6$ & 8.0 & 6.9\\	
%\hline
%\end{tabular}
%\end{center}
%\end{table}

\subsection{Large radial excursion of marginally stable systems.}
In general, the binary systems and the fragments formed in unstable
circumbinary rings have non circular orbits. Thus, most
brown dwarfs formed close to the binary are likely to be 
ejected.  But, brown dwarfs' with initial semi major axis $a_b \sim a_c$,
can be scattered to large distances from the center of mass of the
system without escaping from its gravitational potential within the
timescale $T_d$
(although it might still escape after all).
We call these cases ``marginally stable'' because they could be unstable
if one could integrate for larger timescale or 
use non-numerical methods to prove the instability rigorously.

In Fig. 3, 
We illustrate three such
examples each with $a_b \sim a_c$.  In model 1, we choose $\mu = 0.1$,
$e_\ast = 0.4$, $a_b = 4.4$, $e_b =0$ and the argument of brown
dwarf's apogee, $\theta_b=90^o$.  In this case, the brown dwarf
reaches to 200 binary separation by the end of simulation,
i.e. $t=T_d$.  In model 2, ($\mu = 0.1$, $e_\ast =0.6$, $a_b = 7.8$,
$e_b =0.4$, and $\theta_b=180^o$), the brown dwarf's orbit expands to
50 times binary's initial separation at $t \sim 0.6 T_d$.   But, 
subsequently at $t=
T_d$, the extent of the brown dwarf's radial excursion
is reduced to approximately its initial value.  In
model 3, ($\mu = 0.5$, $e_\ast =0.2$, $a_b = 6.3$, $e_b =0.4$, and
$\theta_b=270^o$), the excursion reaches to 100 initial binary
separation.  These examples indicate that the existence of wide and
marginally stable orbits and that {\it under some marginal circumstances, brown
dwarfs can be scattered to large distance from but remain bound to some
main-sequence binary stars for a particular time scale.}
% The two brown dwarf at distances $??$ AU
%from the young binary star GG Tau (ref ?? ) and 
The discovery
of a brown dwarf candidate at a distance of $\sim 100$ AU from a young binary
star, TWA 5
%TWA 6 Hyd 
may be examples of such a system (Lowrance et al. 1999).
%The brown dwarfs still could escape in the future and the system GG Tauri
%could be % (McIntosh, 2000?).

\subsection{Ejection speed of escapers} 
Brown dwarfs with $a_b < a_c$ are ejected from the gravitational 
potential of 
their host binary system.  We now examine their escape speed. For this
study, we first consider the case with $\mu=0.5$ and $e_\ast = 0$.  We
approximate the brown dwarf companions as massless particles with $e_b
= 0$ initially.  Because brown dwarfs are assumed to be test
particles and thus they do not interact each other, 
we can put more particles in one simulation
to investigate different initial conditions and get better statistics.
 We place 24 particles around a ring with a radius
$a=1.3$ from the center of mass of the binary.  These particles are
initially separated in the azimuthal direction by 15$^o$ between any
two closest pairs. An additional 24
particles with similar azimuthal spacing are placed on a circle with
$a=1.5$.  All of these particles have $a < a_c$ such that all of them are
ejected from the proximity of the binary's orbit within $10^4$ binary
periods.

We also carried another series of calculations with $\mu = 0.5$ and
$e_\ast=0.2$ where 24 particles are placed on each of two concentric
rings with $a=1.9$ and 2 respectively.  Their azimuthal positions are
equally separated by $15^o$.  Finally, for the $e_\ast = 0.4$ and 0.6
case, we placed 24 particles in a similar manner on each of two
concentric rings with $a=2.4$ and 2.5.  Thus, for each binary star, we
place 48 particles with $a<a_c$ around the binary. All of these
particles are ejected from the system.

We summarize all the results in Fig. 4 in which we plot the
distribution of the escape speed. These results show  that the
ejection speed is typically half the orbital speed of the binary.  
%In \S2, we showed that fragmentation of the circumbinary disk occurs when
%its most unstable region is optically thin.  The condition for
%gravitational instability is $Q = c_s \Omega / \pi G \Sigma \sim 1$
%whereas the condition for the disk to be optically thin is $\kappa
%\Sigma < 1$.  In the outer regions of protostellar disks where ice
%grains dominate, the opacity $\kappa = \kappa_0 T^2$ where $\kappa_0 =
%2 \times 10^{-4}$ cm$^2$ / gm Kelvin$^2$.  Hydrostatic equilibrium in
%the vertical direction implies that the thickness of the disk $H = c_s
%/ \Omega$.  From these three relations, we find that the disk can be
%both gravitationally unstable and optically thin only outside a
%critical radius $R_b \simeq (\kappa_0 G^2 M_\ast^3 / \pi R_g^2)^{1/4}
%(H/R)^{5/4} \sim 3 \times 10^{16} (H/R)^{5/4} $ cm.  (Recall
%that these are the two necessary conditions for disk fragmentation.)
%If the disk is
%heated by the stellar irradiation, the aspect ratio $ H/R \sim 0.1 ??$
%(Hayashi {\it et al.} 1985) and $R_b \sim 3 \times 10^{15}$ cm.  
%The discussions in \S2 suggest that rings with sufficiently high
%surface densities at $R_b$ may be attained by gas infall onto
%the tidal barrier of a binary system with a separation
%$a$ slightly less than $R_b$. 
For $R_b \sim 3 \times 10^{15}$ cm,
$a \sim R_b$, and the total mass of binary system to be $\sim 1 M_\odot$,
the binary's orbital speed would need to be
$\sim 3-5$ km s$^{-1}$.  Our results  show that the
escaping speed of the brown dwarf ejecta is $\sim 1-3$ km s$^{-1}$.
In a young stellar cluster, such as the Orion, this ejection speed is
a fraction of the velocity dispersion of the cluster which is in a
dynamical equilibrium.  Thus, {\it brown dwarfs ejected from the close
proximity of the binary would not generally escape the gravitational
potential of the cluster.}  This result is consistent with the large
concentration of brown dwarfs in young stellar clusters such as the
Orion complex (Lucas \& Roche 2000).

\section{Formation of brown dwarf pairs}
%The results in \S2 indicate that the growing fragments in a unstable
%circumbinary disk perturb each other's motion.  During the early
%growth stage, close encounters may induce energy dissipation which
%enables the fragments to capture each other.  The captured fragments
%may either coalesce or form binaries with separation less than the
%Roche radius of their combined masses.  Close binary brown dwarfs may also
%be formed through secondary fragmentation since each of the fragment
%also has a considerable amount of angular momentum.  Such hierarchial
%fragmentation of isothermal rotating cloud has been simulated
%numerically (Bodenheimer ???).  
Indeed, close brown dwarf binaries
have been found (Koerner et al. 1999), but these systems are generally
not orbiting around some other binary main sequence stars. 
%(Note that
%brown dwarf binaries formed this way should have nearly identical
%metallicities). 
Similar to single brown dwarfs, close binary brown
dwarfs may also be strongly perturbed by the gravity of the binary and
be ejected.

%In this section we consider the following issues: 1) if close binary brown
%dwarfs were formed through the fragmentation of unstable circumbinary
%disks, would they be able to survive their close encounters with their
%host binary stars? and 2) if fragmentation of the circumbinary disk
%resulted in the formation of a collection of single brown dwarfs, what
%is the probability for them to capture one of their siblings while they
%are being ejected from their host binary environment?

\subsection{Survival of pairs}
In order to test the survival probability of the brown dwarf binaries,
we first place a massless test particle to simulate the dynamics of a
secondary companion around the brown dwarf (For some interesting
cases, a series of models with masses for both brown dwarf companions
are included).
%, see below for details).  
This approximation allows us to
first explore the range of parameters which may be favorable for the
survival of the brown dwarf pairs.  Based on the results in Fig.1 and Fig. 2,
we can identify the range of model parameters which leads to ejection
of brown dwarf fragments.  As a test, we adopt $\mu = 0.5$ for the
binary star and choose $\mu_b=0.05$, $a_b = 2.3$ and apogee at
$0^\circ$ for the primary of
the brown dwarf binary.  The secondary of the brown dwarf binary is
assigned with an initial semi major axis $2.5$ and apogee at $0^\circ$, thus
the separation of the brown-dwarf pair is $0.2$ which is inside the
Roche radius of the primary ($R_R = (\mu_b/3)^{1/3} = 0.25$).  We also
assume that the center of mass of the brown-dwarf pair is initially on
a circular orbit around the center of mass of the binary system and the
brown-dwarf secondary is on a circular orbit around the brown-dwarf
primary.

Then, we try four cases of different eccentricities of central binary
stars relative to each other: $e_\ast=0.0, 0.2, 0.4, 0.6$.  
%The orbital evolution of this
%system is integrated for $10^3$ central binary periods.  
Our results
indicate that the brown-dwarf pairs would survive their ejection from
the neighborhood of the binary in the low-eccentricity 
($e_\ast=0.0$ and $e_\ast=0.2$) limit.  But, in the limit that the 
binary system has a large eccentricity (i.e. for the
$e_\ast=0.4$ and $e_\ast=0.6$ models), the brown dwarf pairs have
a tendency to become dissociated.
%(Figure 3-6.) 
Therefore,
it is plausible that brown dwarf pairs may remain bound to each other 
during their ejection from the binary system's gravitational potential.
But, it is also possible for their ejection to produce two freely 
floating single brown dwarfs.

The above approximation provides a useful tool for us to identify the
range of parameters which allows a brown dwarf pairs to survive.  The
massless approximation for the secondary is applicable to brown dwarf
binary with extreme mass ratios.  We now takes the next iteration by
replacing the ($\mu_b = 0.05$) primary and a massless secondary
brown-dwarf pair with a system of two equal-mass ($\mu_b = 0.025$)
brown dwarf companions.  We find, with identical four sets of model
parameters as above, brown dwarf binaries  all 
remain intact as they are
ejected by their host binary stars for these four cases 
$e_\ast=$0.0, 0.2, 0.4 and 0.6.

We also enlarge the separation of the brown-dwarf binary to 0.3 which
is larger than its Roche radius.  Again all four sets of initial
conditions are used.  In all these four cases, 
the brown dwarf binary is
always disrupted during its close encounters with the binary star.
%Thus, {\it brown dwarf binaries with separations less than their Roche
%radius are likely to survive the ejection from their host binary star}
%because the ejection of the brown dwarf does not involve close
%encounter with either of their two stellar components.
%In contrast, when their separation is larger than the Roche
%radius of the brown dwarf pair, the binding force between brown dwarfs 
%is weaker than the tidal force from the central binary so that
%the brown dwarf pairs are easily
%disrupted.
 
\subsection{Pair capture}

We now explore the possibility that both brown dwarfs were formed as 
single brown dwarfs and that they may have captured each other to become
a binary.
In order to evaluate this probability, we repeat the earlier simulation
in which 24 particles are placed in a ring around the binary system
(see Section 2.3).
The main difference with the earlier models is that a mass of $\mu_b
= 0.05$ is assigned to each particles.  The corresponding Roche radius
for each individual particle is $\sim 0.25 $
 which is comparable to
their initial separation. 
All the particles are ejected from the
gravitational potential of the binary system but no particle captured
any other particle.  
%To make sure that this conclusion is correct, we even put 24
%formation region. 
%%%We did not see any case that the capture happened.
We thus conclude this second scenario is very unlikely.

%The possibility that brown dwarf pairs formed in isolation faces the
%same difficulty that their individual and total masses are much smaller
%than the characteristic Jeans mass in typical molecular clouds.  If they
%were formed through fragmentation, they must be removed from the vicinity
%of rich gas supply. Finally, we did not investigate the possibility of 
%tidal capture between two freely floating brown dwarfs which may be the 
%relic of ejection around other binary systems.  But the probability of
%such a process may be too small to account for the frequency of close 
%short-period brown dwarf pair.

\section{Concluding Remarks}
%Our present numerical scheme does not adequate resolution to 
%resolve the collapse and evolution of these fragment into the full
%nonlinear limit. Interestingly, some of these fragments have approximately
%axisymmetric structure about their local center of mass while others 
%appear to be elongated.  The former fragments may collapse into single
%brown dwarfs.  But, the latter fragments appear to be rapid rotators 
%and they are tidally perturbed by the binary system.  Because they 
%are optically thin, we expect them to contract further and undergo 
%fragmentation and form binary or multiple gravitationally bound 
%substellar-mass objects.
We have proposed that large number of brown dwarfs 
are probably formed through fragmentation
in circumbinary disks. We study the criteria for ejection from their
cradles, the possibility to do large radial excursion for
marginally stable systems and also the ejection speed of escapers.  
In addition to that, we also study the formation of brown-dwarf pairs.

For single brown dwarf satellites around binary systems,
our dynamical calculations show that when they are formed
in dynamically unstable regions, they are likely to be
ejected from the gravitational potential of the binary system.
These results provide an explanation for 
%the common sighting of
the discovery of field brown dwarfs (Ruiz et al. 1997, Tinney 1998)
and also consistent with the discovery of brown dwarfs as cluster members
(Rebolo et al. 1995, Lucas \& Roche 2000).

% and brown dwarf pairs (ref..)
%}

For binary brown dwarf satellite pairs, the calculations of four body (main
sequence binary with brown-dwarf pair) interaction show that these 
brown-dwarf pairs can remain to be bound to each other during the 
ejection if their initial separation is well within their Roche radius
(which is typically $R_R \sim 0.2-0.25$ binary separation). 
Thus, {\it brown dwarf pairs with separations less than their Roche
radius are likely to survive the ejection from their host binary star}
because the ejection of the brown dwarf does not involve close
encounter with either of their two stellar components.
In contrast, when their separation is larger than the Roche
radius of the brown dwarf pair, the binding force between brown dwarfs 
is weaker than the tidal force from the central binary so that
the brown dwarf pairs are easily
disrupted. 
These results might help to investigate the 
formation histories of brown-dwarf pairs detected
in Koerner (1999) and Basri \& Martin (1999).
% in which the projected 
%physical separations for pairs are between 5 and 10 AU only.}

%.....
%In addition to brown dwarf, our theory can also find application 
%to the problem of L,T, and M dwarfs.
%observational predictions ??
We thank S. Aarseth, P. Bodenheimer, M. Duncan, E. Martin,
M. Bate, M. Sterzik and B. Reipurth
for useful conversations. We are grateful to the referee's good suggestions.
This research has been supported in part by the 
National Science Council, Taiwan, under Grants NSC 91-2112-M-008-018
and, in the U.S. part, by the
NSF through grants AST-9618548, AST-9714275, by NASA through grants
NAG5-4277 and NAG5-7515, and an astrophysics theory program which
supports a joint Center for Star Formation Studies at NASA-Ames
Research Center, UC Berkeley, and UC Santa Cruz.

\clearpage
\section*{REFERENCE}
\begin{reference}

\reference Armitage, P. J., Bonnell, I. A. 2002, MNRAS, 330, L11

\reference Aarseth, S. J., Lin, D. N. C. \& Palmer, P. L. 1993, \apj, 403, 351

\reference Basri, G., \& Martin, E. L. 1999, \apj. 118, 2460

\reference  Bate, M. R., Bonnell, I. A., Bromm, V. 2002, MNRAS, 332, L65

\reference Beckwith, S. V. W.,  Sargent, A. I., Chini, R. S.
           \&  Guesten, R. 1990, AJ, 99, 924

\reference Burkert, A., \& Bodenheimer, P. 1996, MNRAS, 280, 1190

\reference Cochran, W. D., \& Hatzes, A. P. 1994, Astrophys. Space Sci., 
212, 281

%\reference Dvorak, 2003, this volume

\reference Epchtein, N., et al. 1994, in Science with Astronomical
Near-Infrared Sky Surveys, ed. N. Epchtein, A. Omont, B. Burton, \& P. Persei
(Dordrecht: Kluwer), 3

\reference Gizis, J. E., Kirkpatrick, J. D., Burgasser, A., Reid, I. N., 
Monet, D. G., Liebert, J. \& Wilson, J. C. 2001, ApJ, 551, L163

\reference Gizis, J. E., Reid, I. N., Knapp, G. R., Liebert, J., 
Kirkpatrick, J. D., Koerner, D. W. \& Burgasser, A. J. 2003, AJ, 125, 3302 

\reference Henry, T. J. \& McCarthy, D. W. 1993, \aj, 106, 773

\reference Holman, M. J. \& Wiegert, P. A. 1999, \aj, 117, 621 

\reference  Koerner, D. W., Kirkpatrick, J. D., McElwain, M. W. 
\& Bonaventura, N. R. 1999, \apj, 526, L25 

\reference Lin, D. N. C.,  Laughlin, G., 
              Bodenheimer, P., Rozyczka, M. 1998, Science,281, 2025

\reference Lowrance, P. J. et al. 1999, ApJ, 512, L69

\reference Lucas, P. W., Roche, P. F. 2000, MNRAS, 314, 858

\reference Makino, J. \& Aarseth, S., J. 1992, PASJ, 44, 141 

\reference Marcy, G. W. \& Benitz, K. J. 1989, \apj 344 441

\reference Mayor, M. \& Queloz, D. 1995, Nature, 378, 355

\reference Nakajima, T. Oppenheimer, B. R., Kulkarni, S. R., Golimowski, D. A.,
Matthews, K. \& Durrance, S. T. 1995, Nature, 378, 463

\reference Neuh\"auser, R., Guenther, E. W., Petr, M. G., Brandner, W.,
           Hu\'elamo, N., Alves, J. 2000, A\&A, 360, L39

\reference Oppenheimer, B. R., Golimowski, D. A., Kulkarni, S. R., 
           Matthews, K., Nakajima, T., Creech-Eakman, M. 2001, 
           AJ, 121, 2189 

\reference Rebolo, R., Zapatero-Osorio, M. R., Martin, E. L. 1995, Nature, 377,
           129

%\reference Reid, I. N., Kirkpatrick, J. D., Liebert, J., Burrows, A.,
%Gizis, J. E., Burgasser, A., Dahn, C. C., Monet, D., Cutri, R.,
%Beichman, C. A. \& Skrutskie, M. 1999 \apj, 521, 613

\reference  Reipurth, B., Clarke, C. J. 2001, AJ, 122, 432

\reference Roddier, C., Roddier, F., Northcott, M. J., Graves, J. E., 
           Jim, K. 1996, ApJ, 463, 326

\reference Ruiz, M. T., Leggett, S. K., \& Allard, F. 1997, \apj, 491, L107

\reference Skrutskie, M. F., et al. 1997, in The Impact of Large-Scale 
Near-IR Sky Survey, ed. F. Garzon et al. (Dordrecht: Kluwer), 187

\reference Tinney, C. G. 1998, MNRAS, 296, L42
\reference Walker, G. A. H., Walker, A. R., Irwin, A. W., Larson, A. M.,
Yang, S. L., \& Richardson, D. C. 1995, Icarus, 116, 359

\end{reference}

\clearpage

\begin{figure}[tbhp]
\epsfysize 7.0in \epsffile{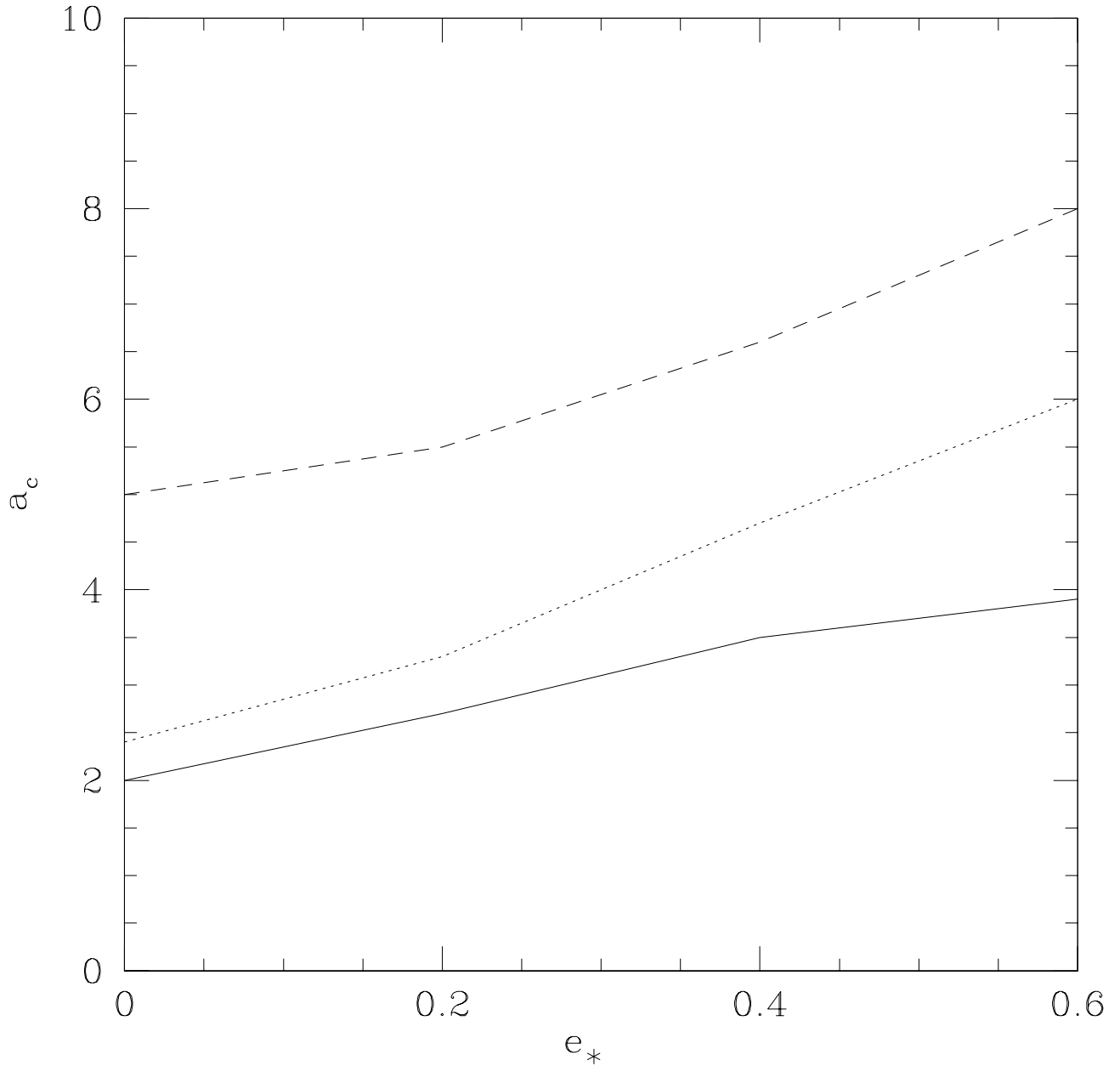}
\caption{The critical semi major axis as function of eccentricity
when the mass ratio of the binary system $\mu=0.1$.
The simulations were run at $e_{\ast}=$ 0.0, 0.2, 0.4 and 0.6 
and then the results connected by lines.
The solid line represents the result for models with brown dwarfs been
treated as massless test particles which initially move on circular orbits.
%These models are analogeous to those obtained by  
%Holman \& Wiegert (1999). 
The dotted
line represents the results for models in which  brown dwarfs 
are assigned with finite mass $\mu_b= 0.05$
% The brown dwarfs are 
and assumed to have  circular orbits
% about the center of mass of the binary system 
initially.  The dash line represents the results of models in which 
brown dwarfs have mass $\mu_b=0.05$ and have initial eccentricity
$e_b=0.4$. 
%(See Table 2). 
}
\end{figure}

\clearpage

\begin{figure}[tbhp]
\epsfysize 7.0in \epsffile{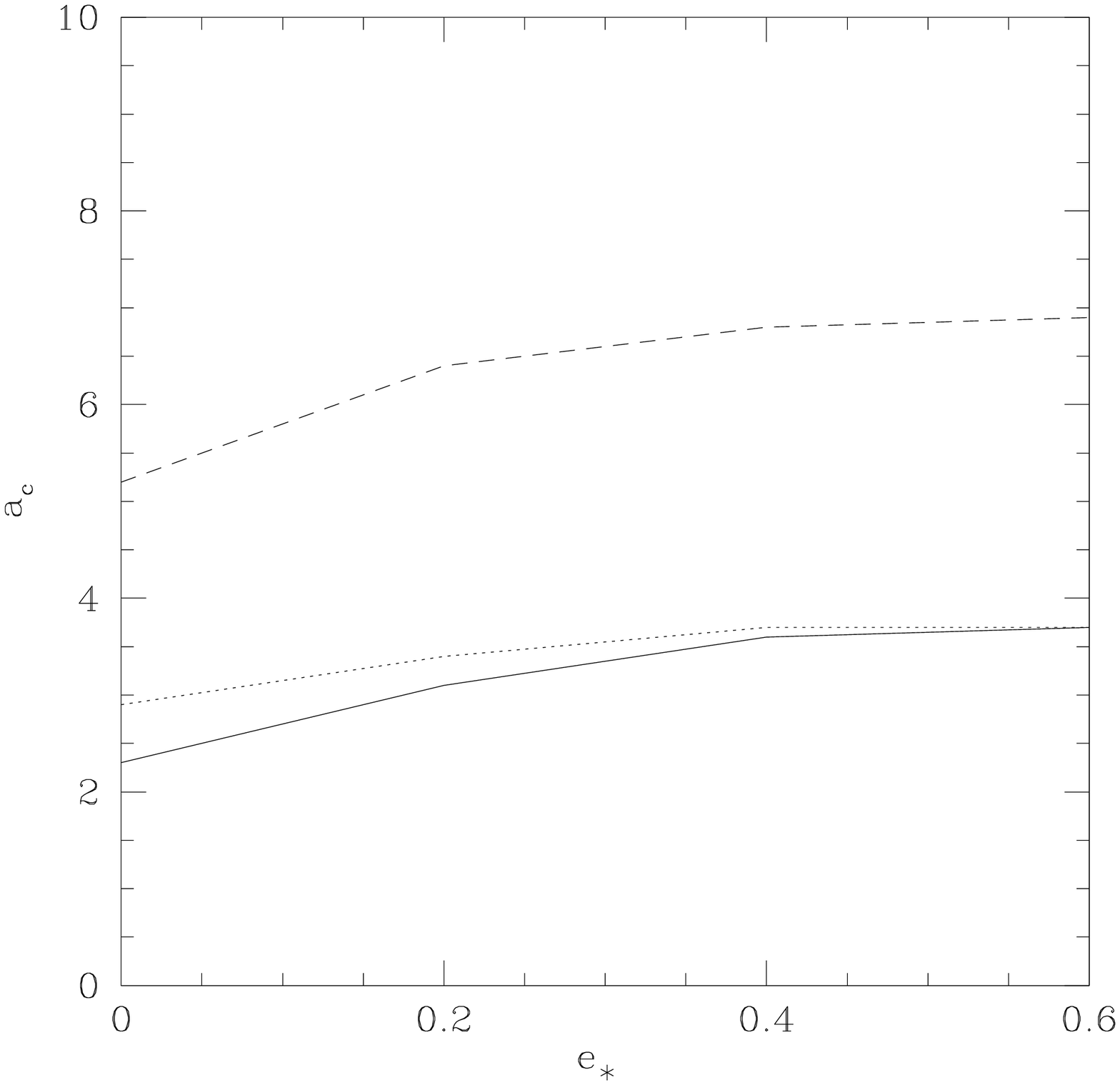}
\caption{The critical semi major axis as function of eccentricity
when the mass ratio of the binary system $\mu=0.5$. 
The simulations were run at $e_{\ast}=$ 0.0, 0.2, 0.4 and 0.6 
and then the results connected by lines.
The solid line represents the result for models with brown dwarfs been
treated as massless test particles which initially move on circular orbits.
%These models are analogeous to those obtained by  
%Holman \& Wiegert (1999). 
The dotted
line represents the results for models in which  brown dwarfs 
are assigned with finite mass $\mu_b= 0.05$ 
%The brown dwarfs are 
and assumed to have  circular orbits 
%about the center of mass of the binary system 
initially.  The dash line represents the results of models in which 
brown dwarfs have mass $\mu_b=0.05$ and have initial eccentricity
$e_b=0.4$.
%(See Table 2). 
}
\end{figure}

\clearpage

\begin{figure}[tbhp]
\epsfysize 7.0in \epsffile{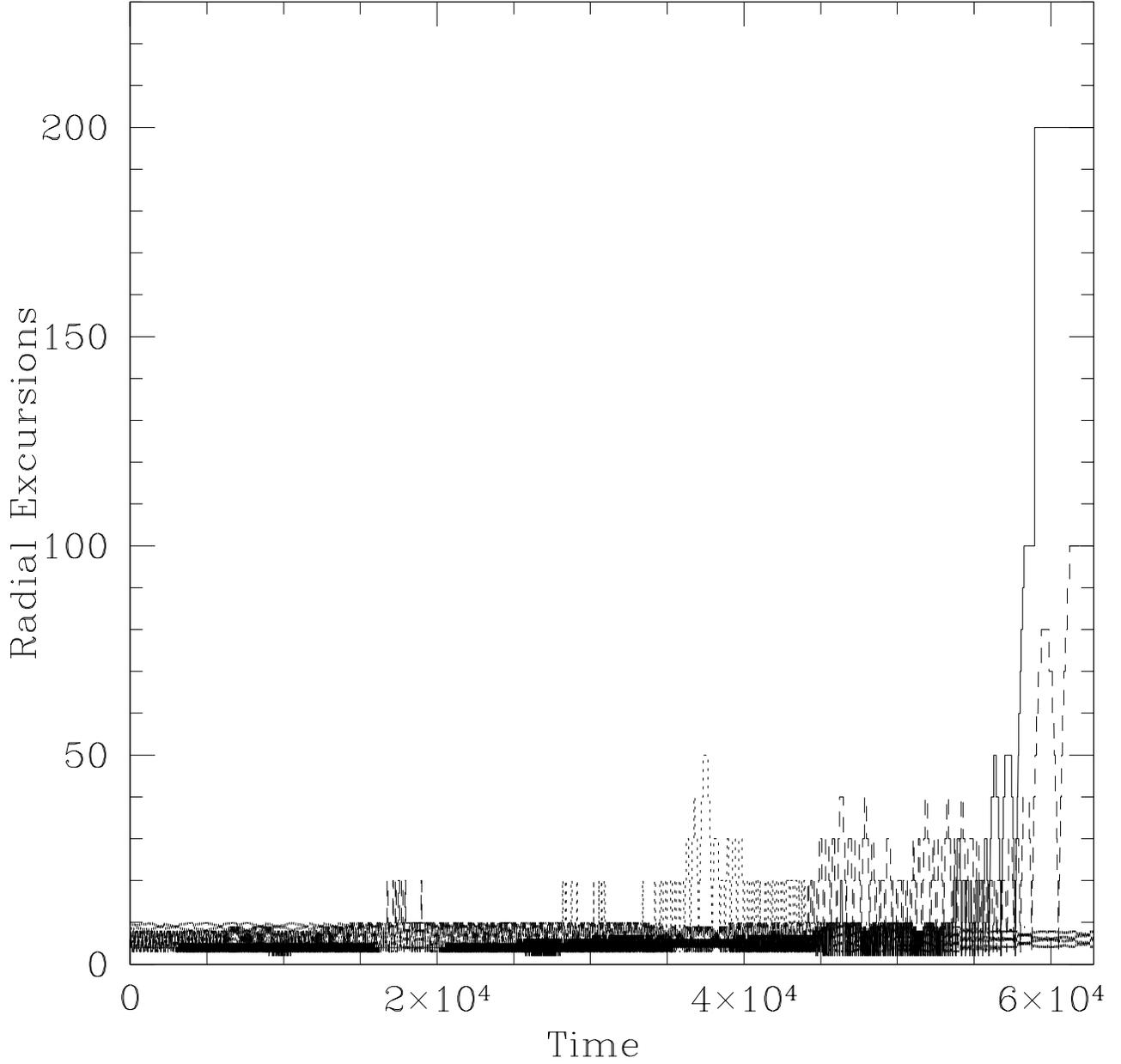}
\caption{The radial excursions as function of time for model 1-3 in Section
2.2, where the solid line represents the results for model 1. 
The dotted and dash lines represent the results for models 2 and 3 
respectively.
}
\end{figure}

\clearpage

\begin{figure}[tbhp]
\epsfysize 7.0in \epsffile{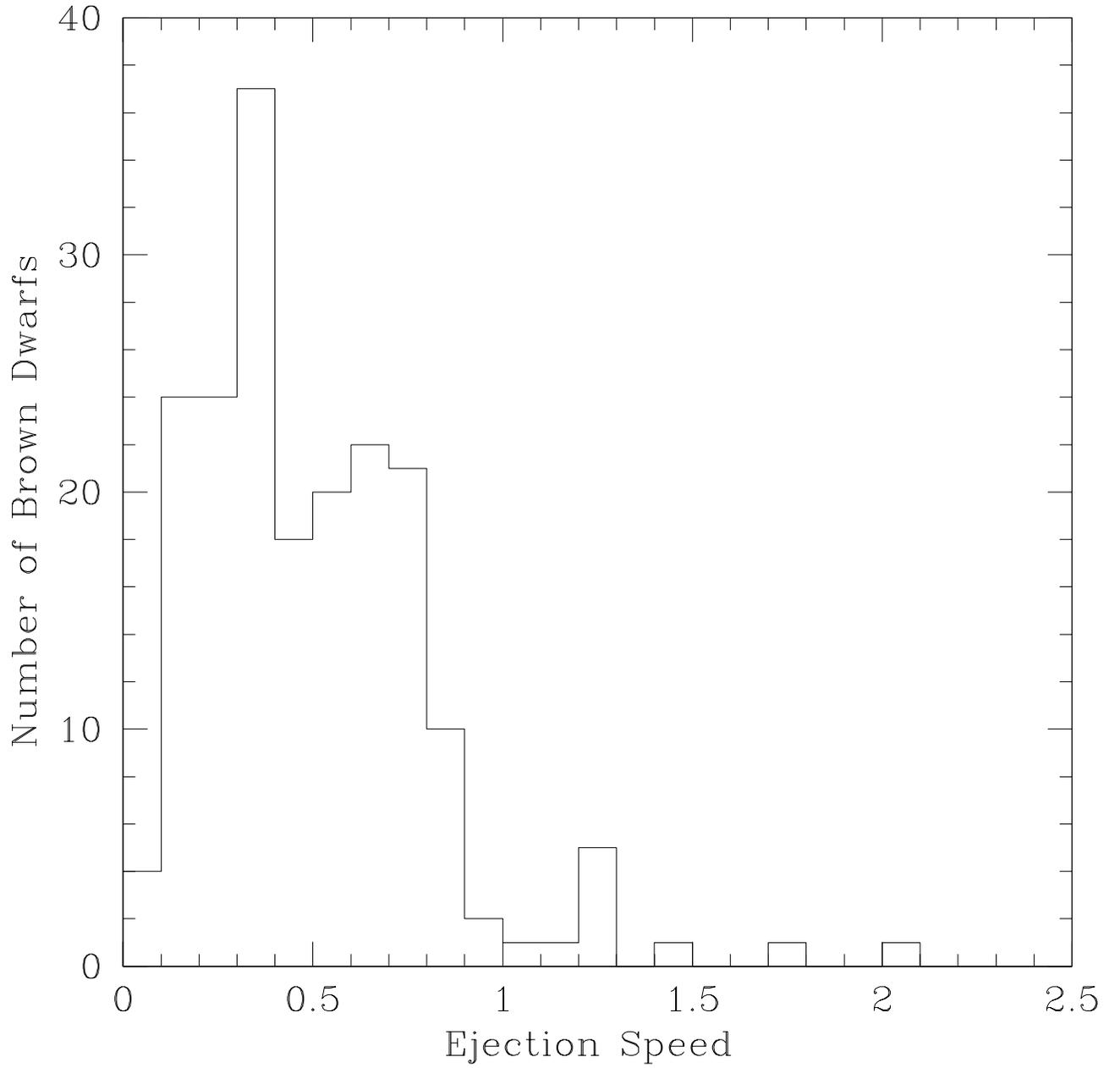}
\caption{The histogram for the ejection speed of brown dwarfs,
where the binary's orbital velocity is 1.}
\end{figure}

\end{document}